\documentclass[psfig,epsf]{aipproc}
\layoutstyle{6x9}
\begin{document}
\newcommand{\mkeq}[1]{\begin{equation}#1\end{equation}}
\newcommand{\firstphi}{\phi}
\input epsf
\input psfig.sty

\title{
Critical Phenomena Associated with Boson Stars
}

\author{Scott H. Hawley}{
  address={Max Planck Institut f\"ur Gravitationsphysik, 
           Albert Einstein Institut,
           14476 Golm, Germany},
  email={shawley@aei.mpg.de} 
}
\author{Matthew W. Choptuik}{
 address={ CIAR Cosmology and Gravity Program, 
         Department of Physics and Astronomy, 
         University of British Columbia, Vancouver, British Columbia, 
         Canada V6T 1Z1 },
 email={choptuik@physics.ubc.ca},
}

\maketitle

\vspace{-0.7cm}
{\small \it 
Here we present a synopsis of related work \cite{HawleyChop,Hawley}
describing a study of black hole threshold}\hfil\break

\vspace{-.6cm}
\noindent{\small \it phenomena for a
self-gravitating massive complex scalar field in spherical symmetry.}

Studies of models of gravitational collapse have revealed structure which
can arise near the threshold of black hole formation.  The solutions in
this regime are known as ``critical solutions," and their properties as
``critical phenomena''.  These solutions can arise generically, even
in simple models such a massless scalar field in spherical symmetry
\cite{Mattcrit}.

Critical solutions can be constructed dynamically via numerical
simulations, in which one considers continuous one-parameter families 
of initial
data with  the following ``interpolating'' property: for sufficiently
large values of the family parameter, $p$, the evolved data describes
a spacetime containing a black hole, whereas for sufficiently small
values of $p$, the matter-energy in the spacetime disperses to large
radii at late times, and {\em no} black hole forms.  Within this range
of parameters, there will exist a critical parameter value, $p=p^\star$,
which demarks the onset, or threshold, of black hole formation.

Over the past decade, numerical and closed-form studies of collapse in
various matter models have enlarged the picture of critical phenomena
\cite{Dave,MattYM,Brady,AE}, so that we now have a more complete
understanding of the relevant dynamics.  (Interested readers should
consult the reviews by Gundlach \cite{CarstenRev,CarstenLivRev} for a more
comprehensive discussion of critical phenomena.)  Black-hole-threshold
solutions are {\it attractors} in the sense that they are almost
completely independent of the specifics of the particular family used as
a generator.  Up to the current time, the only initial data dependence
which has been observed in critical collapse occurs in models for
which there is more than one distinct black-hole-threshold solution.
Critical solutions are by construction unstable, having precisely
one unstable mode \cite{EC,Koike}. Thus letting $p\rightarrow
p^{\star}$ amounts to minimizing or ``tuning away'' the initial amplitude
of the unstable mode present in the system.  These solutions also possess
additional symmetry which, to date,  has either been a time-translation
symmetry, in which the critical solution is static or periodic, or
a scale-translation symmetry (hometheticity), in which the critical
solution is either continuously or discretely self-similar (CSS or DSS).

These symmetries are indicative of the two principal types of critical
behavior that have been seen in black hole threshold studies (with some
models exhibiting both types of behavior depending on the initial data).
For Type I solutions, there is a finite minimum black hole mass which can
be formed, and there exists a scaling law for the lifetime $\tau$ of the
near-critical solutions such that $\tau \sim -\gamma \ln|p-p^{\star}|$
where $\gamma$ is a model-specific exponent.  In Type II critical
behavior, a black hole of arbitrarily small mass can be formed, and the
critical solutions are generically {\em self-similar}.

Our current interest is a critical-phenomena-inspired study of the 
dynamics associated with ``boson stars'' 
\cite{Kaup,RB,Colpi,Jetzer,Mielke}.  A boson star is given by a complex
massive scalar field $\firstphi(t,r) =  \phi_0(r) \exp (i\omega t)$,
minimally coupled to gravity as given by general relativity.  In this
study, we dynamically construct critical solutions of the Einstein
equations coupled to a massive, {\em complex} scalar field $\phi(t,r)$,
by simulating the implosion of a spherical shell of {\em massless}
real scalar field $\phi_3(t,r)$ around an ``enclosed'' boson star.
The massless pulse then passes through the origin, explodes and continues
to $r\rightarrow \infty$, while the massive complex (boson star) field
is compressed into a state which ultimately either forms a black hole or
disperses.  For the massless field $\phi_3(0,r)$, we choose a gaussian
of fixed width $\Delta$ and initial distance $r_0$ from the origin,
and vary the amplitude $A$ until the critical solution is obtained
(to within machine precision).

\begin{figure}
\epsfxsize = 11cm
\epsfysize = 10cm
\centerline{\epsffile{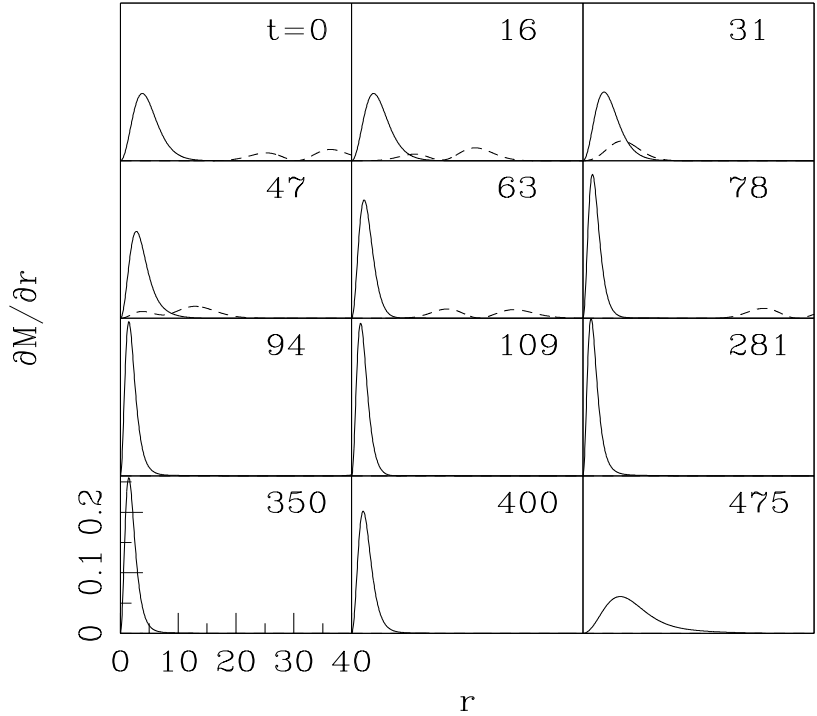}}
\caption{
Evolution of a perturbed stable boson star with $\phi_0(0)=0.04\times\sqrt{4\pi}$
and mass $M_C=0.59 M_{Pl}^2/m$.  This shows contributions to the radial derivative
of the total mass $\partial
M(t,r)/\partial r$ due to the massive field $\phi(t,r)$ (solid line) and massless 
field $\phi_3(t,r)$
(dashed line).  We start with a stable boson star centered at the origin,
and a gaussian pulse of massless field.  (We see two peaks for
the massless field because it is only the gradients of $\phi_3$,
not $\phi_3$ itself, which contribute to $\partial M/\partial r$.)  In the 
evolution
shown above, the pulse of massless field enters the region containing
the bulk of the boson star ($t\simeq 15$), implodes through the origin
($t \simeq 30$) and leaves the region of the boson star ($t \simeq 50$).
Shortly after the massless pulse passes through the origin, the boson star
collapses into a more compact configuration, about which it oscillates
for a long time before either forming a black hole or dispersing.
(The case of dispersal is shown here.)
} \label{fig:anim}
\vspace{-1cm}
\end{figure}

\begin{figure}
\centerline{ \psfig{figure=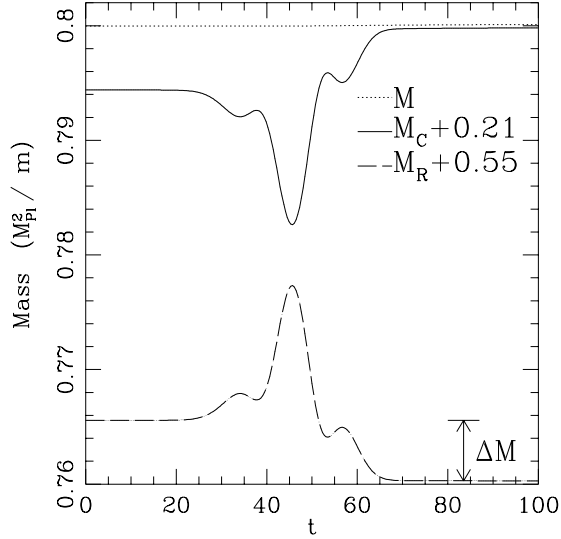,height=7.5cm,width=7.5cm} }  
\caption{
Exchange of energy between the real and complex scalar fields.  The solid
line shows the mass of the complex field, shifted upward by
$0.21 M_{Pl}^2/m$ for graphing purposes.  The long-dashed line shows the 
mass of the real field,
shifted upward by $0.55 M_{Pl}^2/m$.  The amount (and percentage) of mass
transfer goes to zero as we consider boson star initial data approaching
the maximum mass (the transition to instability).  The dotted line near
the top of the graph shows the total mass $M = M_C+M_R$, which is 
conserved to within a few hundredths of a percent.
In all cases, we only see a net transfer of mass {\it from} the real
field {\it to} to complex field, and not vice versa.
}
\label{fig:massexchange}
\end{figure}

Figure \ref{fig:anim} shows a series of snapshots from a typical
simulation in which the parameter $p$ ($p \equiv A$) is slightly below
the critical value $p^{\star}$, for an initial stable boson star with
a mass of $M = 0.59 M_{Pl}^2/m$ (where $M_{Pl}$ is the Planck mass).
In this figure, we have plotted the individual contributions
of the complex and real fields to the total mass
of the spacetime.  That is, we have defined masses $M_C$ and $M_R$ of the
complex and real fields, respectively.  (Only in vacuum regions and for
times at which the supports of the two fields do not overlap can
$M_C$ and $M_R$ be interpreted as physically meaningful masses.)
During this gravitational ``collision,'' mass is
is transferred from the real to the complex field, as shown in Figure
\ref{fig:massexchange}.

The resulting critical solutions persist for some finite amount of time
which depends on the fine-tuning $p-p^\star$ of the initial data.  As we
have shown in \cite{HawleyChop}, the lifetimes $\tau$ of the near-critical
solutions follow the scaling law for Type I solutions, $\tau = -\gamma
\ln|p-p^\star|$.  Here $\gamma$ is related to the imaginary part of the
growth factor $\sigma$ of the unstable mode ($\sim \exp[i\sigma t]$)
by $\Im(\sigma)=1/\gamma$.  In keeping with the Type I nature of these
solutions, we find a finite minimum mass for the black hole formed 
for as we let $p\rightarrow p^\star$ (for $p>p^\star$).

\begin{figure}
\centerline{ \psfig{figure=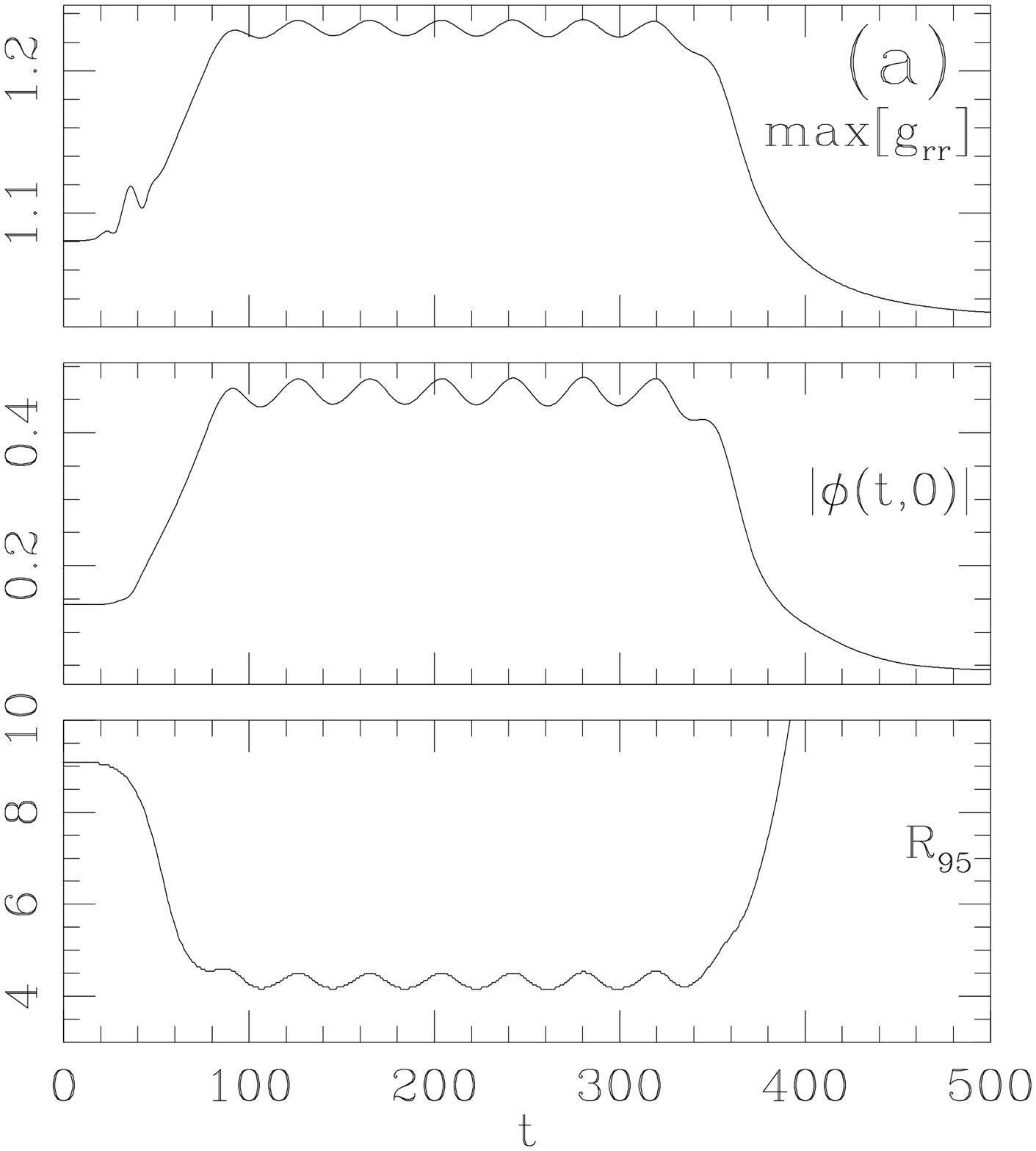,height=8.0cm,width=7.5cm}
             \psfig{figure=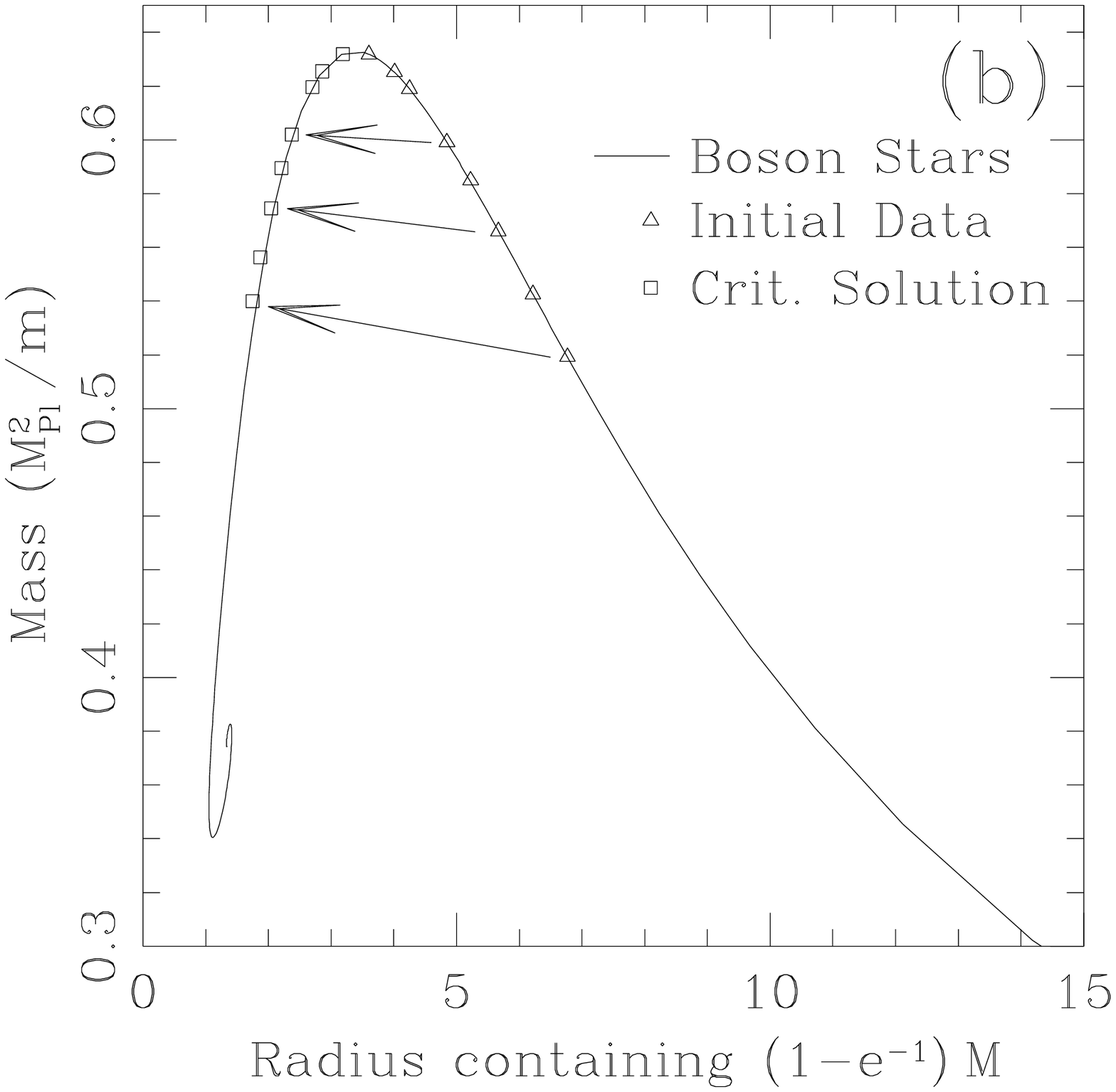,height=8.3cm,width=7.5cm}}
\caption{ 
(a) Quantities describing a near-critical solution.  Here we show
timelike slices through the data shown in Figure 1, an evolution which
ends in dispersal.  Top: Maximum value of the metric function 
$g_{rr}(t,r)$.  Middle: Central value $|\phi(t,0)|$ of the massive field.
Bottom: Radius $R_{95}$ containing 95\% of the mass-energy in the complex
field.  (b) Mass {\it vs.} radius for equilibrium configurations of boson
stars (solid line), initial data for the complex field (triangles), and
critical solutions (squares).  Arrows are given to help match initial
data with the resulting critical solutions.  Points on the solid line
to the left of the maximum mass $M_{\rm max}\simeq 0.633 M_{Pl}^2/m$
correspond to unstable boson stars, whereas those to the right of the
maximum correspond to stable stars.  The squares show
the time average of each critical solution, which exists during the
oscillatory regime shown in (a).  We show the radius containing
$(1-e^{-1})M\simeq 0.63M$ instead of $R_{95}$ in order to exclude the
halo which forms in the critical solution (see Figure \ref{fig:haloanim}).
}
\label{fig:mvr63} 
\end{figure}

\begin{figure}
\centerline{
	\hbox{\psfig{figure=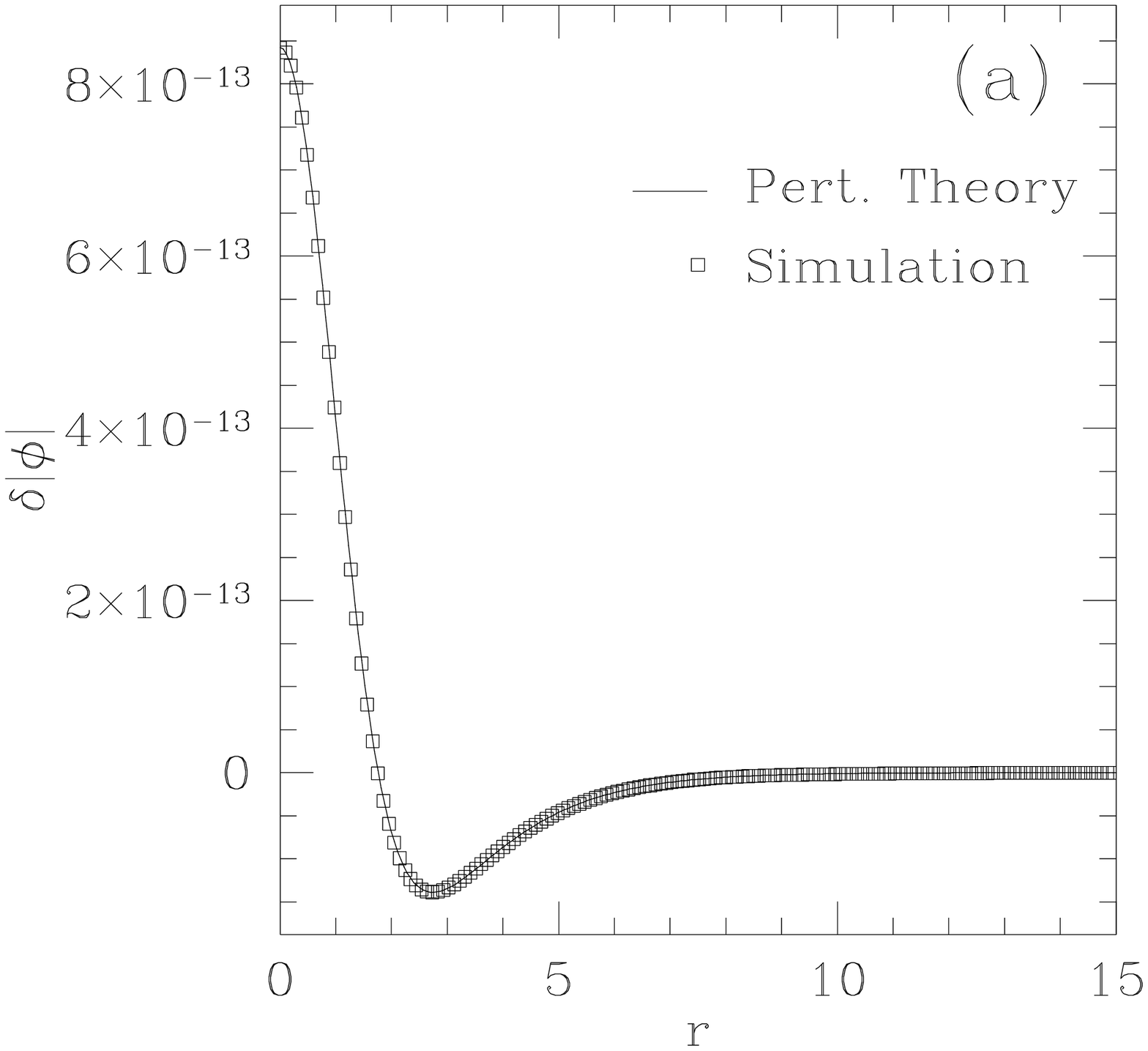,height=8.5cm,width=7.9cm}} 
	\hbox{\psfig{figure=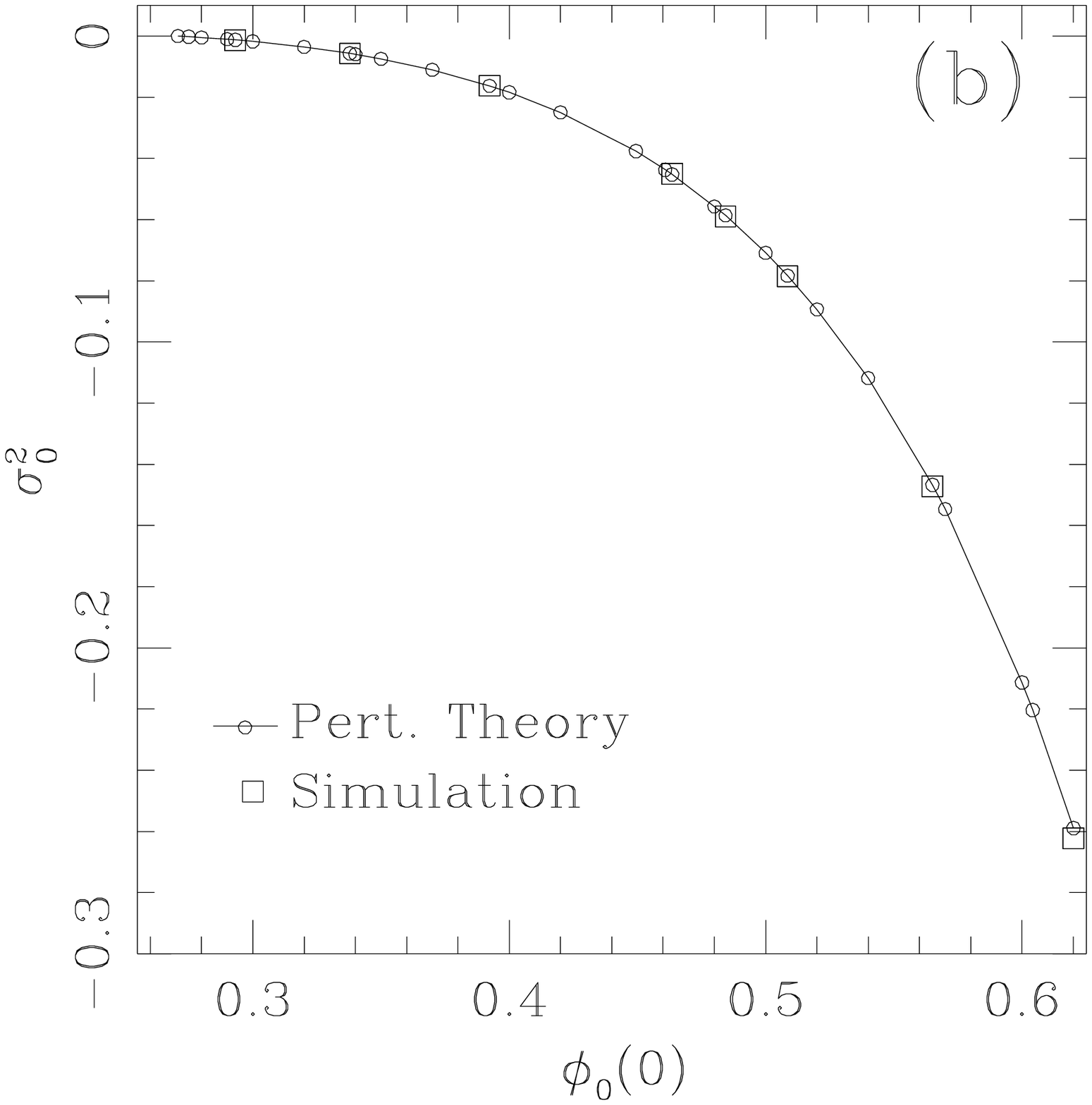,height=8.5cm,width=7.5cm}} 
}
\caption{
(a) Fundamental mode of unstable boson star.  The solid line shows
$\delta |\phi|$ obtained from the perturbation theory calculations.
The squares show the difference between $|\phi|$ for two simulations for
which the critical parameter $p$ differs by $10^{-14}$.  Differences
between the simulation data and perturbation theory results are below
$1.1\times10^{-15}$.  (b) Comparison of squared growth factors 
(squared Lyapunov exponents) $\sigma_0^2$ for unstable
modes.  The circles show a subset of the perturbation theory results we
obtained for unstable boson stars.  The squares show the measurements
of unstable growth factors in our simulations.  (The solid line simply
connects the circles.)
}
\label{fig:unstab_phi} 
\end{figure}

The critical solutions have properties which correspond closely with those
of unstable boson stars, as shown in Figure \ref{fig:mvr63}.  To further
extend the comparison between these critical solutions and boson stars,
we perform a linear perturbation analysis about boson star equilibria,
building on the work of Gleiser and Watkins \cite{GW}.  Using the method
described in \cite{HawleyChop}, we find the distribution for the squared
frequency $\sigma^2$ of boson star quasinormal modes with respect to
$\phi_0(0)$, and we find the radial shapes of the modes.  In Figure
\ref{fig:unstab_phi} we provide a comparison between unstable modes found
from our simulations and corresponding results obtained via perturbation
theory about a boson star which is a "best fit" to the simulation data.
We find close agreement between the shapes and frequencies of the unstable
modes obtained by these two different methods.  A comparison of the next
higher (oscillatory) mode also yields favorable results.

Thus the critical solutions we obtain appear to correspond to boson
stars exhibiting superpositions of stable and unstable modes.
For boson stars with masses somewhat less than the maximum boson star mass
$M_{\rm max}\simeq 0.633 M_{Pl}^2/m$ ({\it e.g.} those boson stars with
masses $0.9 M_{\rm max}$ or less), however, we find less than complete
agreement between the critical solutions and unstable boson stars of
comparable mass.  This is evidenced by the presence of an additional
spherical shell or ``halo'' of matter in the critical solution, located
in what would be the tail of the corresponding boson star.  Interior to
this halo, we find that the critical solution compares favorably with the
boson star profile.  

\begin{figure}
\centerline{
        \hbox{\psfig{figure=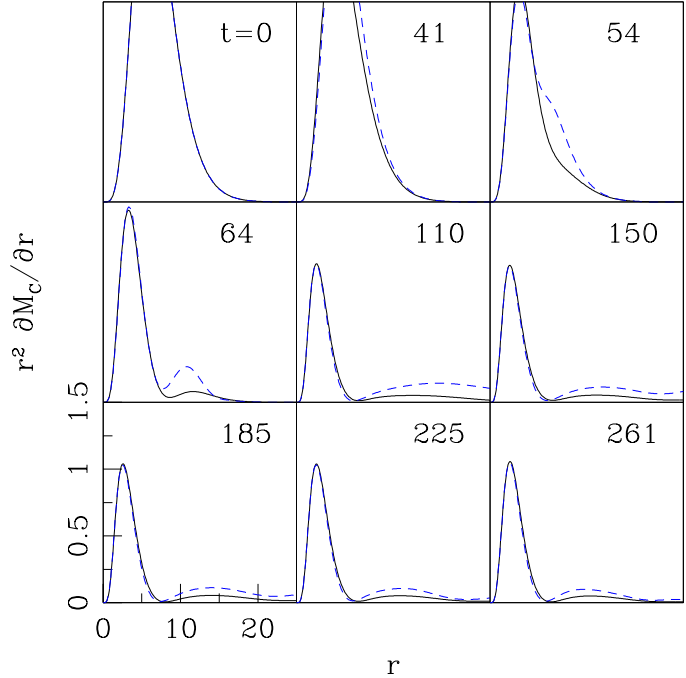,height=9.2cm,width=9.2cm}}
}
\caption{
Evolution of $r^2 \partial M_C/\partial r$ for for two different sets of initial data.
Both sets contain the same initial boson star, but the massless field
$\phi_3(0,r)$ (not shown) for one set is given by a gaussian whereas for the other
set $\phi_3(0,r)$ is given by a ``kink" .  The amplitude of each pulse is
varied (independently for each family of initial data) to obtain the critical 
solution.
We have multiplied $\partial M_C/\partial r$ by $r^2$ to highlight the dynamics of the
halo; thus the main body of the solution appears to decrease in size as
it moves to lower values of $r$.  The kink data produces a larger and
much more dynamical halo, but interior to the halo, the two critical
solutions match closely --- and also match the profile of an unstable
boson star.  Thus, the portion of the solution which is ``universal''
corresponds to an unstable boson star.  One can see the additional halo of
matter in the region roughly $8\leq r \leq 23$, exterior to the bulk
of the critical solution.
}
\label{fig:haloanim}
\end{figure}

It is our contention that this halo is not part of
the true critical solution, but is instead an artifact of the collision
with the massless field.  As one might expect, the properties of the
halo are not universal, {\it i.e.} they are quite dependent on the type
of initial data used.  In contrast, the critical solution interior to
the halo is largely independent of the form of the initial data.
To demonstrate this, we use two families of initial data, given
by a gaussian a ``kink" ($\phi_3(0,r) = A/2 [1+\tanh[(r-r_0)/\Delta]$).
A series of snapshots from one such pair of evolutions is shown in
Figure \ref{fig:haloanim}.  We suspect that the halo is radiated over
time (via scalar radiation, or ``gravitational cooling'' \cite{SandSGC})
for all critical solutions.  We find, however, that the time scale for
radiation of the halo is comparable to the time scale for dispersal or
black hole formation for each (nearly) critical solution we consider.
With higher numerical precision, one might be able to more finely tune
out the unstable mode, allowing more time to observe the behavior of
the halo before dispersal or black hole formation occur.

For the future, we consider it worthwhile to investigate similar scenarios
for neutron stars. While there have bee studies regarding the explosion
of neutron stars near the minimum mass ({\it e.g.,}\cite{ENS1}), we
would like to see whether neutron stars of {\it non-minimal mass} can
be driven to explode via dispersal from a critical solution.


\vspace{3cm}
\begin{thebibliography}{00}
\vspace{-0.45cm}
\bibitem{HawleyChop}{S.H. Hawley and M.W. Choptuik, Phys. Rev. D {\bf 62}, 104024 (2000).}
\bibitem{Hawley}{S.H. Hawley, Ph.D. Dissertation, University of Texas at Austin (2000).}
\bibitem{Mattcrit}{M.W. Choptuik, Phys. Rev. Lett. {\bf 70}, 9 (1993).}
\bibitem{Dave}{D.W. Neilsen and M.W. Choptuik,  Class.Quant.Grav. {\bf 17}, 
   761-782 (2000).}
\bibitem{MattYM}{M.W. Choptuik, T. Chmaj and P. Bizon, Phys. Rev. Lett.
       {\bf 77}, 424 (1996).}
\bibitem{Brady}{P.R. Brady,  C.M. Chambers and S.M.C.V. Con\c{c}alves,
                       Phys. Rev. D {\bf 56}, 6057 (1997).}
\bibitem{AE}{A.M. Abrahams and C.R. Evans, Phys. Rev. Lett. {\bf 70}, 2980
 (1993).}
\bibitem{CarstenRev}{C. Gundlach, Adv. Theor. Math.Phys. {\bf 2}, 1-49 (199
8).}
\bibitem{CarstenLivRev}{C. Gundlach, Living Reviews 1999-4 (1999).}
\bibitem{EC}{C.R. Evans and J.S. Coleman, Phys. Rev. Lett. {\bf 72}, 1782 (1994).}
\bibitem{Koike}{T. Koike, T. Hara and S. Adachi, Phys Rev. Lett. {\bf 74}, 5170 (1995).}
\bibitem{Kaup}{D.J. Kaup, Phys. Rev. {\bf 172}, 1331 (1968).}
\bibitem{RB}{R. Ruffini and S. Bonnazzola, Phys. Rev. {\bf 187}, 1767 (1969).}
\bibitem{Colpi}{M. Colpi, S.L. Shapiro, and I. Wasserman,
  Phys. Rev. Lett. {\bf 57}, 2485 (1986).}
\bibitem{Jetzer}{P. Jetzer, Phys. Rep. {\bf 220}, 163 (1992).}
\bibitem{Mielke}{E.W. Mielke and F.E. Schunck.  gr-qc/9801063 (1998).}
\bibitem{GW}{M. Gleiser and R. Watkins, Nucl. Phys. B {\bf 319}, 733 (1989).}
\bibitem{SandSGC}{E. Seidel and W.-M. Suen, Phys. Rev. Lett. {\bf 72}, (1994).}
\bibitem{ENS1}{M. Colpi, S.L. Shapiro and S.A. Teukolsky, Astrophys. J. {
\bf 369},  422 (1991).}
\end{thebibliography}
\end{document}